# Evaluating Row Buffer Locality in Future Non-Volatile Main Memories

Justin Meza† meza@cmu.edu
Jing Li* jli@us.ibm.com
Onur Mutlu† onur@cmu.edu

†Carnegie Mellon University *IBM T.J. Watson Research Center

Computer Architecture Lab (CALCM)
Carnegie Mellon University

**Abstract**

DRAM-based main memories have read operations that destroy the read data, and as a result, must buffer large amounts of data on each array access to keep chip costs low. Unfortunately, system-level trends such as increased memory contention in multi-core architectures and data mapping schemes that improve memory parallelism may cause only a small amount of the buffered data to be accessed. This makes buffering large amounts of data on every memory array access energy-inefficient.

Emerging non-volatile memories (NVMs) such as PCM, STT-RAM, and RRAM, however, do not have destructive read operations, opening up opportunities for employing small row buffers without incurring additional area penalty and/or design complexity.

In this work, we discuss architectural changes to enable small row buffers at a low cost in NVMs. We provide a memory access protocol, energy model, and timing model to enable further system-level evaluation. We evaluate the system-level tradeoffs of employing different row buffer sizes in NVM main memories in terms of energy, performance, and endurance, with different data mapping schemes. We find that on a multi-core CMP system, reducing the row buffer size can greatly reduce main memory dynamic energy compared to a DRAM baseline with large row sizes, without greatly affecting endurance, and for some memories, leads to improved performance.

| Date | Notes |
|---|---|
| December 20, 2012 | Initial version of technical report published. |
| September 23, 2013 | Clarified components of row buffer read and write energy. |

Revision notes.

## 1 Introduction

Modern main memory is composed of dynamic random-access memory (DRAM). A DRAM cell stores data as charge on a capacitor. Over time, this charge leaks, causing the stored data to be lost. To prevent this, data stored in DRAM must be periodically read out and rewritten, a process called *refreshing*. In addition, reading data stored in a DRAM cell destroys its state, requiring data to be later restored, leading to increased cell access time and energy. For this reason, DRAM devices require *buffering* data which are read. To keep costs low, the buffering circuitry in DRAM devices is amortized among large *rows* of cells, in peripheral storage called the *row buffer*. Refreshing data and buffering large amounts of data wastes energy in DRAM devices, causing main memory power to constitute a large amount of the total system power (around 40% in some servers [12]).

Data fetched into the row buffer, however, can be accessed at much lower latencies and less energy than accessing the DRAM memory array. Therefore, large row buffer sizes can improve performance and efficiency if many accesses can be served in the same row. Unfortunately, there are several reasons why such *row buffer locality* is low in systems: (1) some applications inherently do not have significant locality within rows (e.g., random access applications), (2) as more cores are placed on chip, applications running on those cores interfere with each other in the row buffers, leading to reduced locality, as also observed by others [26, 24], and (3) interleaving techniques that improve parallelism in the memory system (e.g., cache block interleaving) tend to reduce row buffer locality because they stripe consecutive cache blocks across different banks. As a result, the performance benefit of large row buffers is decreasing in multi-core systems.

New non-volatile memory (NVM) technologies, such as phase-change memory (PCM), spin-transfer torque RAM (STT-RAM), and resistive RAM (RRAM), on the other hand, provide non-destructive reads and do not require refreshing and restoring their data after sensing. This is because NVMs do not store their data as charge, and thus their data persists for a long time and after being read. This not only eliminates the refresh problem of DRAM devices but also opens up opportunities for employing smaller row buffers in NVM's without incurring additional area penalty and/or design complexity.

NVMs such as PCM or STT-RAM are already in or very close to mass production. In 2009, for the first time, a single interface standard was established by JEDEC, called LPDDR2 [2], which allows two types of memory, DRAM and NVM, to share a common bus interface, allowing NVM-based main memories to be built. However, LPDDR2 suffers from two main drawbacks. (1) LPDDR2 does not support banked operation, limiting the amount of parallelism available in the memory system. While this provides low power consumption, it also results in low bandwidth. (2) The "row buffer" as defined in LPDDR2 is different from that used in DRAM based on the JEDEC DDRx standard. Unlike JEDEC DDRx, the LPDDR2 standard specifies two separate data paths for read and write. It does not support a write command to directly access row buffers, instead performing writes through an "overlay window" mapped into the address space of the device. In other words, write data is not written into a row buffer and restored into the array, increasing chip complexity and resulting in a different timing protocol than DDRx SDRAM.

Recently, several NVM prototypes have been reported and demonstrated comparable or even larger capacity than mainstream DDRx SDRAM chips [15, 6, 28]. However, because they are based on the NOR interface [15] or LPDDR2 interface [6, 28], such devices are challenging to build high performance, high bandwidth non-volatile main memories for multi-cores with. In addition, although prior systems work has looked at all-NVM or hybrid DRAM–NVM main memories, such works either assume start-of-art DDRx SDRAM interfaces with only modified technology-related timing parameters or different device organizations entirely [11]. The former suffers from the same problems as DRAM due to its large row buffer assumptions, while the latter incurs high design complexity and area overhead. Moreover, prior work has not addressed the timing protocol changes required to support the proposed implementation for system evaluation.

In this paper, we propose a simple reorganization for NVM banks that can selectively sense and buffer only a small portion of data from NVM array. Since reads are non-destructive in NVM, we swap the column multiplexer (part of the I/O gating circuitry) and the row buffers in the datapath. This allows us to reduce the row buffer size by sharing each basic building block of the new, smaller buffer among multiple columns in the same row. The net result is reduced dynamic energy waste compared to DRAM systems in which we must dedicate buffer resources to all columns in a row with no sharing, even if only a fraction of that data is actually accessed. Unlike LPDDR2, our proposed architecture employs a single datapath for both reads and writes, and allows efficient bank operation to achieve parallelism for high bandwidth. We describe a new memory access protocol for NVM devices to accommodate our new architecture, while retaining a standard pinout and signaling interface (JEDEC DDRx type, identical to DRAM). To enable system-level evaluation, we develop energy and timing models for NVM devices based on their underlying technology parameters.

We evaluate main memory designs with two representative NVM technologies, PCM and STT-RAM, to cover a wide range of NVM characteristics. Our results indicate that small row buffer sizes can provide significant improvements in terms of dynamic main memory energy. Similar to other studies, we find that the contention present on multi-core CMP systems greatly reduces the effect of memory mapping schemes designed to improve locality, reducing the performance benefits of large row buffers on such systems. In addition, all evaluated systems can support reasonable five-year operational lifetimes. We also find that the ability for NVM devices to reduce or eliminate several DDR timings constraints in our new protocol can enable some technologies (such as STT-RAM) to achieve better performance than DRAM.

## 2 Background and Motivation

Emerging NVM technologies have several promising attributes compared to existing memory technologies such as SRAM (used in on-chip caches), DRAM, and Flash. First of all, they overcome the critical drawbacks of legacy NVMs (Flash) such as long access latency and requiring the erasure of large blocks of data during programming. They also provide cost advantages over SRAM and DRAM due to the combination of: (1) small cell size (comparable to or smaller than start-of-art DRAM [18]); (2) simple integration with existing fabrication processes (CMOS); (3) easier cell scalability to smaller sizes (especially compared to DRAM); and (4) available techniques to support multiple bits per cell and/or multiple layers per chip. Importantly, these NVMs feature *non-destructive read operations* which DRAM does not have, (i.e., data sensing does not destroy the contents of cells). However, NVMs also have disadvantages which must be dealt with before they can feasibly be employed as main memory, such as: (1) limited write endurance and (2) high write energy. Even though various techniques have been proposed to extend device lifetime by reducing or balancing the write traffic among all of the available cells in the memory array [11, 17, 7], high write energy is still a serious concern, particularly for NVMs like PCM.

To support our system-level evaluations of these emerging NVM technologies, we parameterize their technology properties as read energy ($\alpha$), write energy ($\beta$), read latency ($\gamma$), and write latency ($\delta$), all normalized to DRAM. In Table 1, we list typical ranges of these parameters based on recent literature surveys [4, 11, 8, 6, 21]. Note that due to a large variety of material options, there is no strict distinction among these technologies electrically, and as such, parameters may vary from material to material even for the same memory technology. In general, however, STT-RAM is faster and has lower read and write energy per cell, and PCM is relatively slow and has much higher write energy per cell. RRAM is the least mature memory, and its characteristics span a broad range across both STT-RAM and PCM. For the remainder of the paper, we divide NVM types into two representative categories: PCM-like and STT-RAM-like.

| Technology | Parameter | PCM | STT-RAM | RRAM |
|---|---|---|---|---|
| Energy | Read ($\alpha$) | 2–8 | 0.5–2 | 0.5–10 |
| | Write ($\beta$) | 10–100 | 1–8 | 1–100 |
| Latency | Read ($\gamma$) | 3–6 | 0.5–2 | 0.5–5 |
| | Write ($\delta$) | 5–30 | 0.5–2 | 0.2–100 |

Table 1: NVM technology parameters, relative to DRAM.

Figure 1(a) shows a typical DRAM chip micro-architecture (JEDEC-standard DDR-type SDRAM). The memory array is divided into banks which consist of rows (wordlines) and columns (bitlines). Due to physical pin limitations, all the information required to service a memory request must be supplied over multiple operations. A memory request begins with a *Row Address Strobe (RAS)* command that sends the row and bank address to select one of the banks and a row within that bank. Then, an entire row is read out and buffered into latch-based sense amplifiers which comprise the row buffer.

The total size of the row buffer is the number of bits read out from a memory array during a bank access (usually 1 to 2KB per chip). Finally, a subset (i.e., column) of data in a row buffer (8B in a DDR3 ×8 device [15]) is selected with a *Column Address Strobe (CAS)* command, processed by I/O gating circuitry, and burst out over the shared memory bus at twice the bus clock rate. Thus, every DRAM access fetches many kilobytes of data and, in the worst case, uses only a tiny portion of it (8B out of 1 or 2KB per chip[1]). If, however, subsequent requests access data in the same row, they can be serviced directly from the row buffer at a reduced energy and latency; this is called a *row buffer hit,* and applications which access data in such a way exhibit high *row buffer locality* (i.e., hit rate in the row buffer). Otherwise, the data in the row buffer has to be restored into the memory array before fetching another row's data in the same bank (called a *row buffer conflict*). The sensing, storage, and restoration of a large amount of data through the row buffer is a major source of wasted energy in existing DRAM architectures.

[1]Note that reading eight ×8 chips comprises an entire 64B cache block.

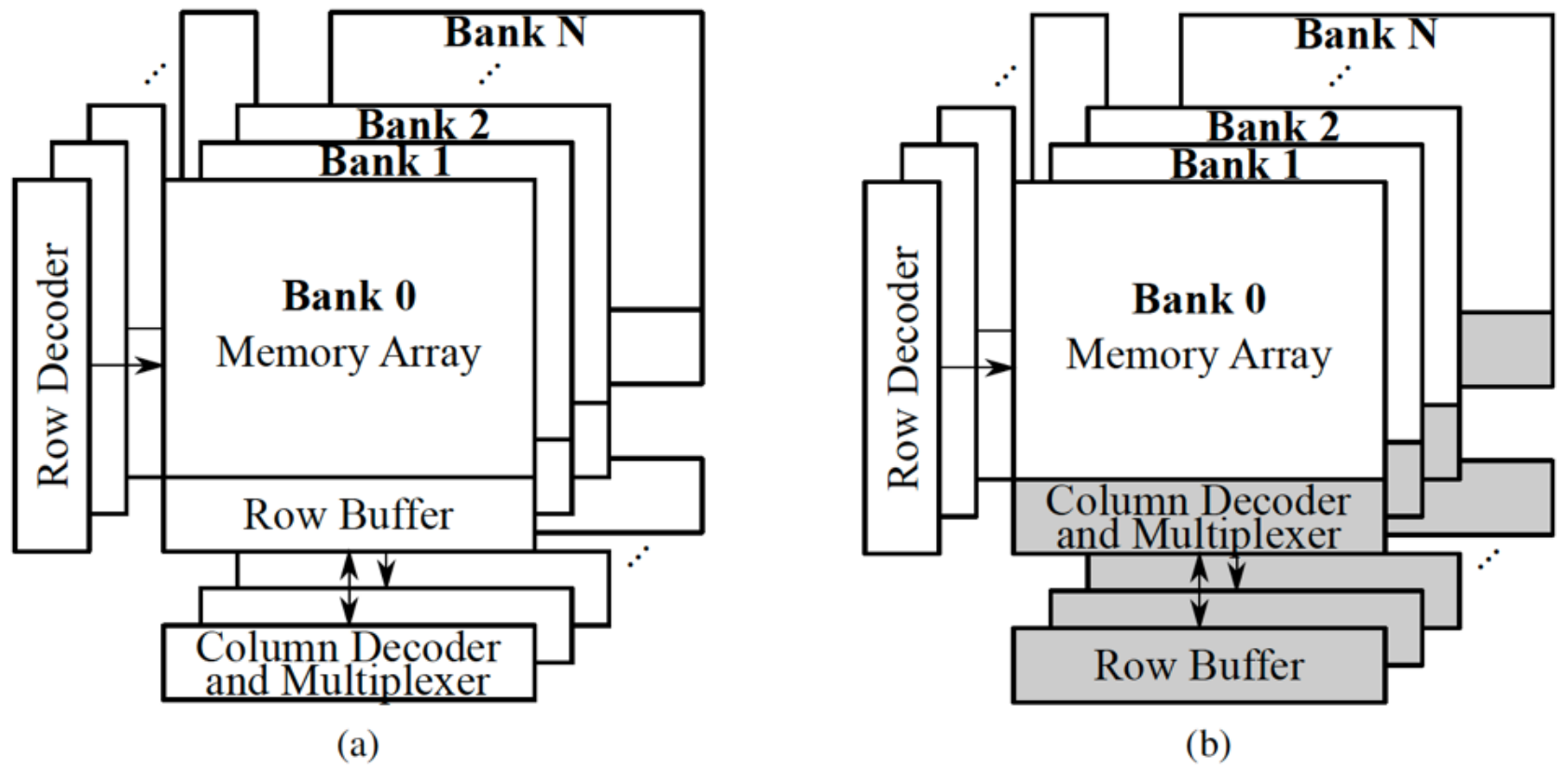

Figure 1: Organization of (a) DRAM and (b) our NVM architecture. The non-destructive read operation of NVM allows the column decoder and multiplexer to be placed before the row buffer, enabling opportunities for reduced row buffer size.

To illustrate how much buffered data is actually used in real applications, Figure 2 shows average row buffer locality when employing various row buffer sizes on several system configurations (we discuss our experimental methodology in Section 4). In particular, we show 1- and 8-core systems employing two different schemes for mapping data in main memory: (1) *row interleaving*, which places consecutive memory addresses in the same row, and (2) *block interleaving*, which stripes data in consecutive memory addresses (usually cache blocks) across different banks. Row interleaving helps exploit row buffer locality by enabling data with spatial locality to reside in the same row buffer, while block interleaving aims to improve memory parallelism by enabling concurrent accesses of memory banks for consecutive memory addresses.

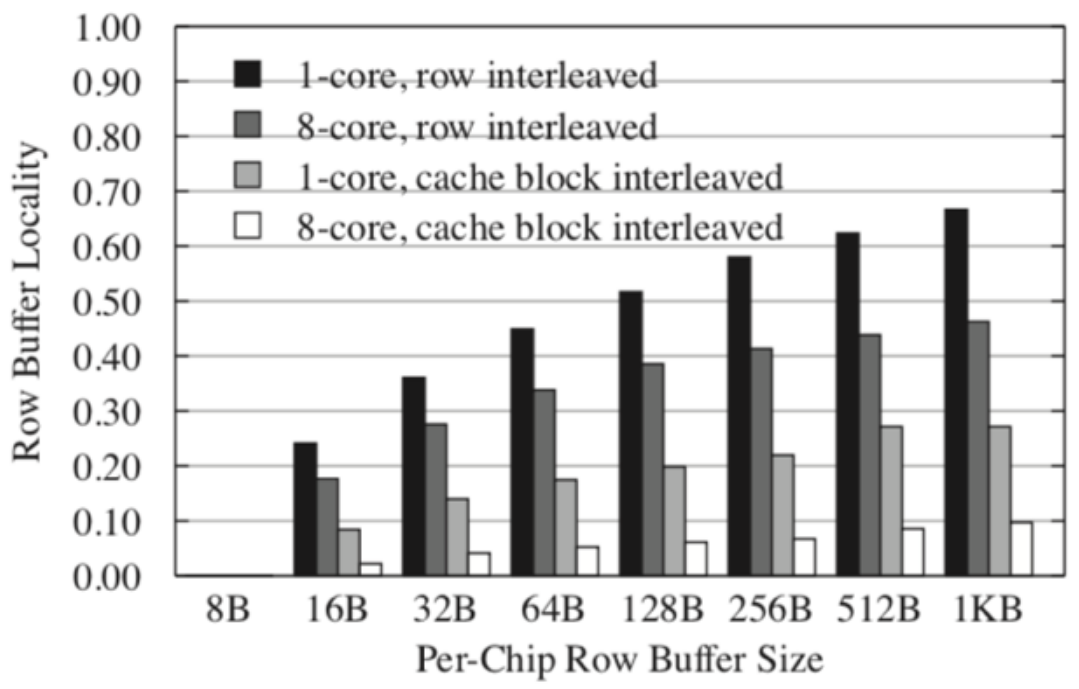

Figure 2: Row buffer hit rates for different row buffer sizes for 1- and 8-core systems with row- and cache block-interleaved address mapping schemes.

We make two observations from Figure 2. First, comparing the 1- and 8-core row-interleaved data, we see that while row interleaving does enable more row buffer locality, its benefits diminish as memory system contention increases with more cores: row buffer hit rate is less than 50% for row interleaving even with large, 1KB rows. Second, block interleaving reduces row buffer locality over row interleaving, to less than 10% in the 8-core case. While it is clear that row locality is decreasing on future many-core CMP systems, what is less obvious is how row buffer size affects system-level tradeoffs, such as energy-efficiency, performance, and durability, in future NVM main memories. This work aims to evaluate these tradeoffs.

# 3 A Small Row Buffer NVM Architecture

Figure 1(b) highlights the key difference in the organization of our NVM architecture compared to DRAM, in which the physical placement of the row buffer and the column multiplexer (part of the I/O gating circuitry in DRAM designs) are swapped in the data path. This rearrangement makes better use of resources by sharing a smaller number of *sense amplifiers* (the devices which store bits in the row buffer) among multiple bitlines. Note that our approach differs from prior NVM device organizations by using the row buffer to provide a single data path for reads and writes for NVM memories, similar to DRAM.

Unlike DRAM, however, our architecture requires decoding both the row address *and* the column address during a *RAS* command, so that only a subset of the row containing the bits of interest will be selected, sensed, and stored in the row buffer. During a *CAS* command, the data bits from the row buffer corresponding to the desired column are further selected by the I/O gating circuitry and sent to a prefetch buffer, as shown in Figure 3. Note that the size of the prefetch buffer is equal to the amount of data fetched per chip to construct a cache block across multiple chips comprising a rank[2]. To accommodate these changes, we propose a new NVM access protocol, based on SDRAM.

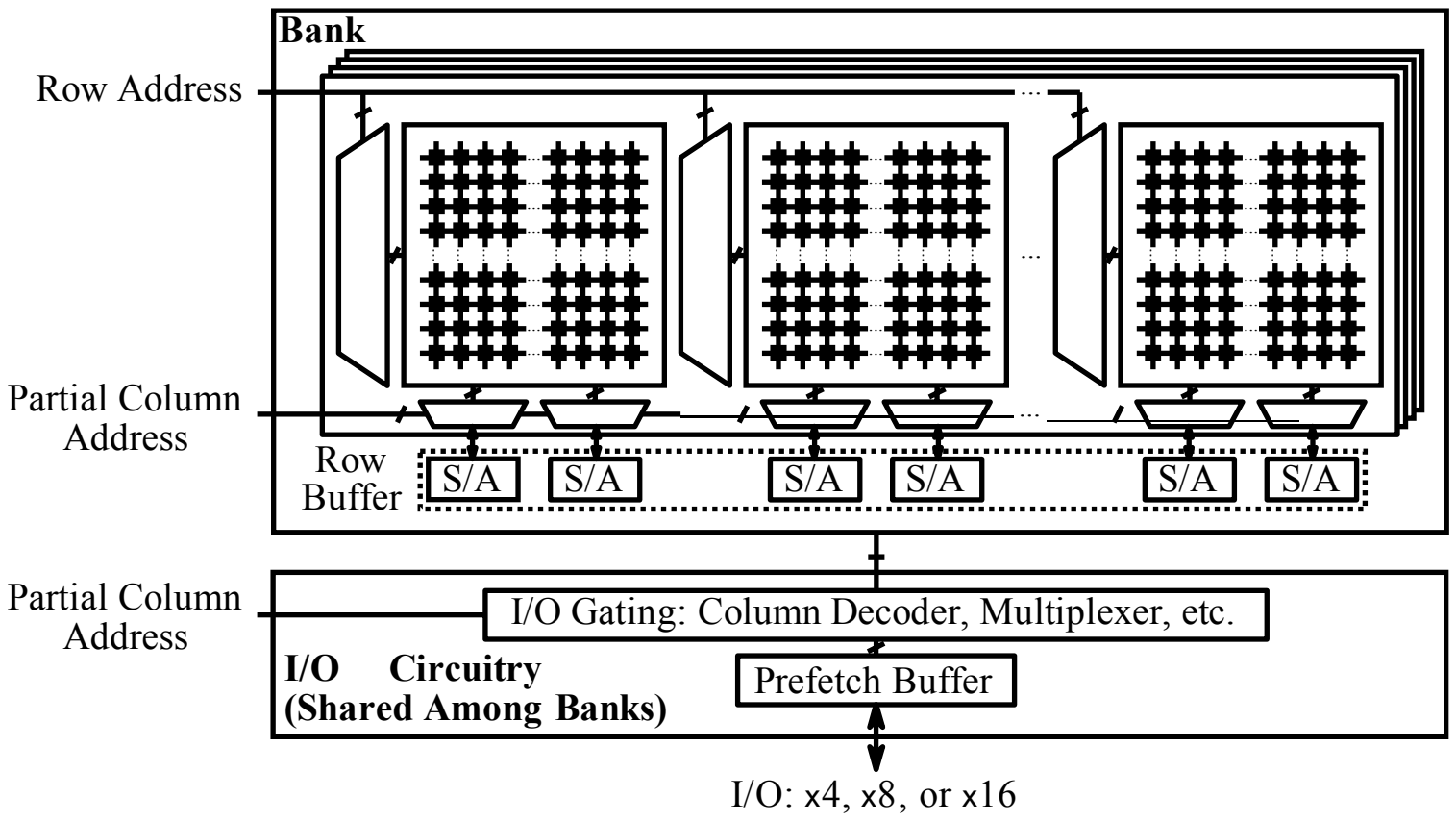

Figure 3: NVM architecture. S/A = sense amplifier.

## 3.1 Memory Access Protocol and Timing Model

The basic DRAM memory access protocol defines commands and timing constraints that a controller uses to manage the movement of data between itself and DRAM chips. There are five basic commands in DRAM: ACTIVATE, PRECHARGE, READ, WRITE, and REFRESH. For our NVM memory access protocol, we attempted to keep the changes compared to the DRAM access protocol at a minimum, while still enabling the key architectural difference of smaller row buffers in NVM. We next compare a DRAM-based protocol to our NVM-based protocol.

### 3.1.1 DRAM Memory-Access Protocol

As described in Section 2, a memory array access in DRAM is completed in two phases due to pin limitations: A *RAS* command (ACTIVATE) is applied to send and decode the row address, perform data sensing, and latch the sensed data into the row buffer; then, a *CAS* command (READ or WRITE) is performed to send and decode the column address, select a subset of data from the row buffer, and transfer them to (during a READ) or from (during a WRITE) the I/O pads. When activating another row in the same bank, the PRECHARGE command evicts the data stored in the latch-based sense amplifiers and resets their state for the subsequent access (this takes $t_{RP}$ time to perform). In standby or various power-down modes, the REFRESH command periodically reads out and restores data in DRAM cells to avoid leakage-induced data loss.

[2]In the special case when the row buffer size is equal to the prefetch buffer size, I/O gating circuitry is not required outside the memory array core.

### 3.1.2 Proposed NVM Memory-Access Protocol

We use four basic commands for NVM operation: ACTIVATE, PRECHARGE, READ, and WRITE. Based on this command set, we use a modified access protocol to address the requirement that both the row address and column address be available before starting an array access.

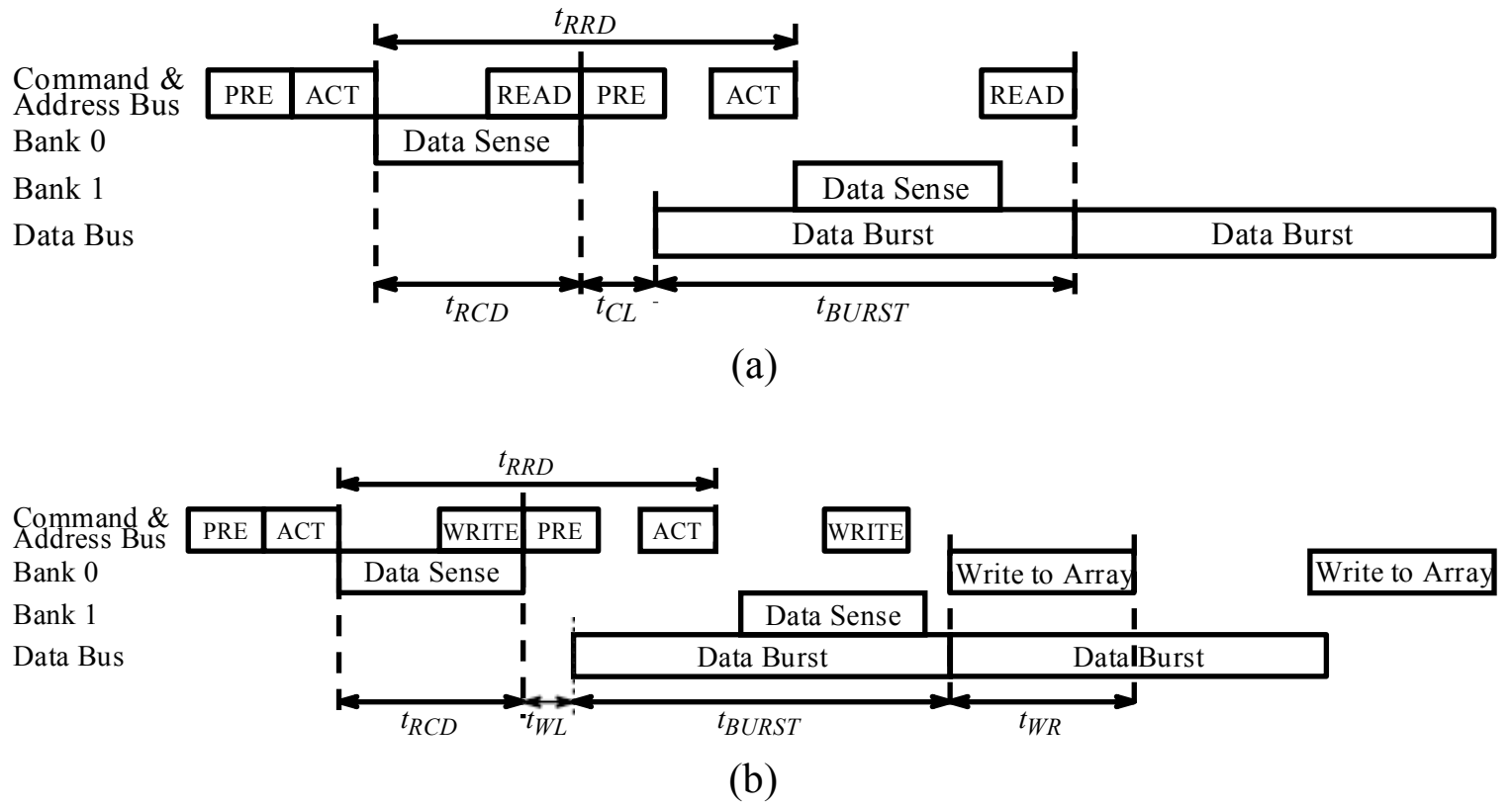

Figure 4: Timing diagram for (a) a read operation and (b) a write operation in our NVM access protocol.

As illustrated in Figures 4(a) and (b), the PRECHARGE command is first performed to send the row address, where it is stored on chip temporarily (similar to the PRECHARGE command defined in LPDDR2 [2]).

Then, the ACTIVATE command is performed to send the column address. Together with the previously-stored row address, the full address bits are decoded to conduct the following operations for array access: (1) selecting the corresponding cells in the memory array, (2) sensing the stored data, and (3) asynchronously loading the sensed data into latches when sensing completes. Unlike LPDDR2, the consequent row buffer access uses a single data path for reads and writes and is therefore much simpler. Depending on whether a READ or WRITE command was issued, the data bits are read from or written into the row buffer and transferred between the row buffer and the I/O's by peripheral logic.

Note that in our NVM design, the PRECHARGE command is not used in the same way as in DRAM. The reason is that in our design, sense amplifiers and latches are kept separately. Therefore, the sense amplifiers are simply sensing units rather than combined sensing and storage units as in DRAM, and can be released as soon as sensing completes. Consequently, the timing required to reset the sense amplifiers can be overlapped with the read or write operation, and hence $t_{RP}$ becomes 0 in NVM. Furthermore, since the PRECHARGE command is simply used to transfer address bits, it is not constrained in its timing ($t_{RTP}$), as in DRAM. We note that this scheme is compatible with various other NVM techniques such as Flip-N-Write [5] to remove redundant write operations for better lifetime and write energy efficiency.

Our timing model for NVM is presented in Table 2, in which we compare it with DDR3-1066 SDRAM [15]. The major timing parameters follow the same naming conventions as in DRAM. Note that $t_{RRD}$ and $t_{FAW}$ specify constraints on the frequency of memory row access between different banks, dictated by peak power constraints. In our evaluated NVM architecture, due to reduced row buffer size, the $t_{RRD}$ and $t_{FAW}$ constraints can be significantly relaxed, meaning more banks can be accessed in parallel and in a much shorter time interval. In addition, all commands can be pipelined when they do not use a shared resource simultaneously. For example, we can initiate multiple bank accesses, one after the other, to achieve a continuous data burst on the bus under the constraints of $t_{RRD}$ and $t_{FAW}$. In fact, if we additionally meet the condition $t_{RRD} \leq t_{BURST}$*, we can achieve close to theoretical peak bandwidth with our protocol.

## 3.2 Energy Model

Despite the different array organizations of DRAM and NVM, the memory chip in each comprises two major sources of power consumption: the memory array core and the periphery. Here, we provide a model which we will use to evaluate the energy of the systems we examine. Equations 1 to 5 show the dynamic energy of DRAM and NVM, where $E$ is dynamic energy, $P$ is dynamic power, $t_{RC} = 1/f$ is cycle time, $V_{DD}$ is internal supply voltage for the

*The total number of cycles to burst data on the bus (4 in this study).

| Parameter | Constraint | DRAM (cycles) | NVM (cycles) |
|---|---|---|---|
| $t_{RCD}$ | A→R/W | 8 | $8\gamma$ |
| $t_{WR}$ | W→P | 8 | $8\delta$ |
| $t_{RRD}$ | A→A | 4 | $4\alpha(m^t/m)$ |
| $t_{FAW}$ | Rate of A | 20 | $20\alpha(m^t/m)$ |
| $t_{RP}$ | P→A | 8 | 0 |
| $t_{RTP}$ | R→P | 4 | 0 |

Table 2: Timing constraints of interest for our study. All other timing constraints are the same as those found in DRAM [15] and clock cycle time is 1.875ns. The command constraints which are enforced are shown using the notation X→Y, where Y shall not be issued less than *timing constraint* cycles after X, and A=ACTIVATE, P=PRECHARGE, R=READ, and W=WRITE. Note that $t_{FAW}$ constrains the maximum number of ACTIVATE commands which can be issued in a time window.

memory core, $V_{DDQ}$ is supply voltage for peripheral circuitry, $m$ and $m^t$ are the size of row buffer in DRAM and NVM, respectively, $n$ is the number of wordlines per bank, $i_{cell,dram}$, $i_{cell,nvm}$ and $i_{leak,dram}$, $i_{leak,nvm}$ are the cell currents for selected cells and the leakage currents for unselected cells in DRAM and NVM, respectively, $I_{DC}$ is the static current of the chip contributed by the on-chip voltage converter and other components, $C_{DE}$ is the output node capacitance of each decoder, and $C_{PT}$ is total capacitance of CMOS logic circuits in the periphery.

$$E \approx P \cdot t_{RC} \tag{1}$$

$$P_{dram} = m i_{cell,dram} V_{DD} + (m \cdot n - m) i_{leak,dram} V_{DD} + (n + m) C_{DE} V_{DD}^2 f + C_{PT} V_{DDQ}^2 f + V_{DD} I_{DC} \tag{2}$$

$$P_{nvm} = m^t i_{cell,nvm} V_{DD} + (m \cdot n - m^t) i_{leak,nvm} V_{DD} + (n + m^t) C_{DE} V_{DD}^2 f + C_{PT} V_{DDQ}^2 f + V_{DD} I_{DC} \tag{3}$$

$$P_{dram} \approx m i_{cell,dram} V_{DD} + C_{PT} V_{DDQ}^2 f \tag{4}$$

$$P_{nvm} \approx m^t i_{cell,nvm} V_{DD} + C_{PT} V_{DDQ}^2 f \tag{5}$$

During reads or writes, the static current $I_{DC}$ and standby current $i_{leak,dram}$, $i_{leak,nvm}$ are negligible compared to the active current $i_{cell,dram}$, $i_{cell,nvm}$. Decoder charging power $[(n + m)C_{DE}V_{DD}^2 f]$ is also small due to the fact that only a small number of nodes are charged at every cycle. Consequently, Equations 2 and 3 can be approximated by Equations 4 and 5. Note that the periphery power required to transfer data between the row buffer and I/O gating circuitry ($C_{PT}V_{DDQ}^2 f$) is the same in both DRAM and NVM, and optimizing this component of power is orthogonal to our study.

The first term in Equations 4 and 5 is the cell activation power required to read data from the array core into the row buffer. In DRAM, reads are performed through RC-based sensing, so $i_{cell,dram} = C_{BL}\Delta V_{DD}$, where $C_{BL}$ is the bitline capacitance and $\Delta V_{DD}$ is the bitline voltage swing. In comparison, NVM usually adopts current-based sensing and, as a result, $i_{cell,nvm}$ is static current and must be kept less than the programming current to avoid overwriting the data being read. It is challenging to reduce sensing current ($i_{cell,nvm}$) or supply voltage ($V_{DD}$), whereas, as we have described, the row buffer size ($m^t$) can be easily reduced in NVM architectures. Consequently, reducing the row buffer size ($m^t$) can achieve significant dynamic energy savings in NVM devices by mitigating cell activation power ($m^t i_{cell,nvm} V_{DD}$).

Based on this analysis, we correlated the energy components in Equations 4 and 5 to the basic memory command set and developed an energy model for NVM, shown in Table 3, in which DRAM energy is also listed for comparison [15].

## 3.3 Area

We estimate the net area impact of our proposed NVM array organization based on the methodology described in [27]. Note that our architecture requires only a reorganization of the row buffer and column multiplexers in the datapath—no additional new circuitry is required. In principle, the net area difference of NVM and DRAM is mainly due to the row buffer area difference, given the same bit cell area (NVM has comparable or smaller bit cell area than DRAM, as described in Section 2).

| Command | DRAM (pJ/bit) | NVM (pJ/bit) |
|---|---|---|
| ACTIVATE | 0.3 | $0.3\alpha$ |
| READ | 19.0 | 19.0 |
| WRITE | 24.2 | 24.2 |
| WRITE$^{t}$ | 0.3 | $0.3\beta$ |

Table 3: Energy model used in our study. ACTIVATE transfers data from the array to the row buffer, READ reads data from the row buffer into the I/O's and drives the I/O's across the bus, WRITE drives data across the bus into the I/O's and writes data from the I/O's into the row buffer, and WRITE$^{t}$ writes data from the row buffer into the array. Peripheral energy is the same for all technologies and is not shown. Note that READ and WRITE include the energy required to drive data across the bus to and from the I/O's, making that component of energy larger than reported in other studies, such as [11].

To provide the same functionality as row buffers in DRAM, the basic building block of NVM row buffers consists of (1) a sense amplifier, (2) a latch, and (3) a write driver. A simple, reconfigurable NVM sense amplifier such as the one proposed in [13] requires only 13 transistors to sense NVM cells which store 1 bit per cell. Taking into account the additional area of a write driver (3T) and an explicit latch (8T), each row buffer in NVM requires 24T, compared to 14T in DRAM [11]. We assume such a device in our study. Note that to achieve better area efficiency than a DRAM-based design, the row buffer area of NVM must be kept approximately less than 58% of that of DRAM. In other words, NVM row buffer size must be less than 512B per chip.

# 4 Methodology

Based on the energy model, timing model, and access protocol presented in Section 3, we evaluated the memory energy and performance characteristics for our NVM-based architecture with smaller rows relative to a baseline DRAM organization across a range of workloads for an 8-core system.

We developed a cycle-accurate DDR3-DRAM memory simulator which we validated against Micron's Verilog behavioral model [16] as well as DRAMSim2 [20]. We use this memory simulator as part of an in-house x86 multi-core simulator, whose front-end is based on Pin [14]. We modify the timings of our memory simulator according to those shown in Table 2 for PCM and STT-RAM. Table 4 shows the major system parameters in the baseline configuration.

| | |
|---|---|
| Processor | 8 cores, 5.3 GHz, 3-wide issue, 8 miss buffers, 128-entry instruction window |
| Last-level cache | 64B cache line, 16-way associative, 512KB private cache-slice per core |
| Memory controller | 64-/64-entry read/write request queues per controller; FR-FCFS scheduler [19] |
| Memory | Timing: DDR3-1066, (timing parameters listed in Table 2). Organization: 2 channels, 1 rank per channel, 8 banks per rank, 8 subarrays per bank, ×8 interface. Block-level writeback of dirty data for NVMs. |

Table 4: Simulation parameters used in our study.

We present results averaged across 31 multiprogrammed workloads composed of 27 benchmarks from SPEC CPU2006, TPC [25] (TPC-C with 64 clients and TPC-H queries 2, 6, and 17), and STREAM [1] (add, copy, and triad). We evaluated 8 high memory intensity (>10 last-level cache misses per kilo instruction) SPEC mixes, 8 medium intensity SPEC mixes (4 high and 4 low intensity benchmarks), 8 low intensity SPEC mixes (no high intensity benchmarks), and 4 TPC server workloads and 3 STREAM workloads both with 8 copies of each benchmark. Each workload ran until all benchmarks in the workload completed at least 100M instructions.

We show results for a DRAM system with different row buffer sizes for comparison, though reducing row buffer size in DRAM incurs significant increased area overhead and design complexity, as discussed in [26, 24]. To compare with NVM-based memories, we will focus on a DRAM chip micro-architecture with 1KB row buffers as our baseline.

Note that in the following discussion, the main memory system is composed of ranks with 8 chips per rank (ignoring additional chips for ECC). To ensure a fair comparison, the NVM- and DRAM-based main memory systems have the same configuration in terms of channels, ranks, and banks. Main memory energy is characterized without including memory controller energy.

# 5 Analysis

As we have seen, row buffer size affects row buffer locality, which can affect system-level metrics. In this section, we analyze the reasons behind the positive or negative impact that row buffer size has on energy, performance, and durability.

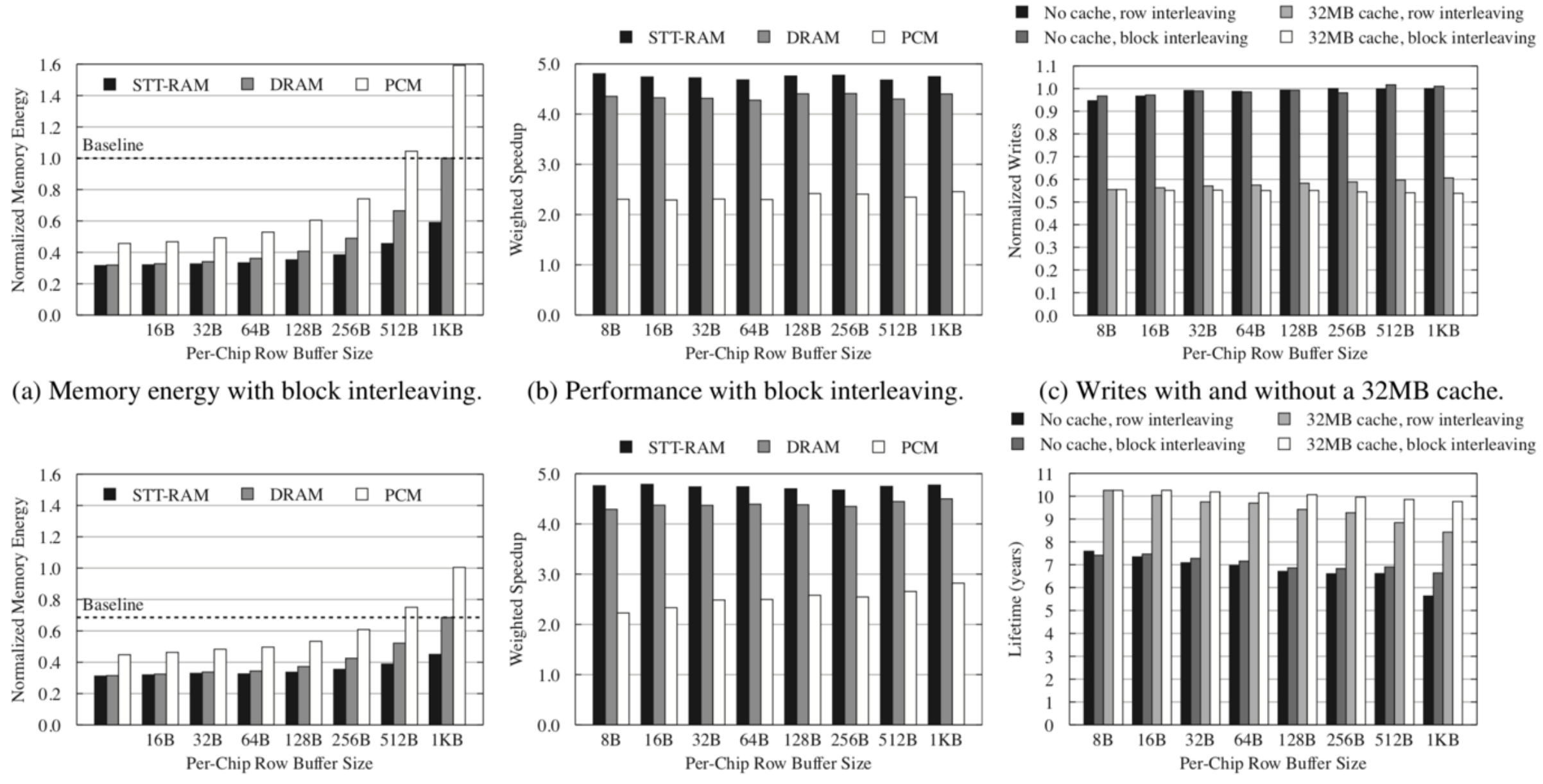

(a) Memory energy with block interleaving. (b) Performance with block interleaving. (c) Writes with and without a 32MB cache.

(d) Memory energy with row interleaving. (e) Performance with row interleaving. (f) Lifetime with and without a 32MB cache.

Figure 5: Multi-core results for energy (left), performance (center), and durability (right). Our baseline is a DRAM system with large 1KB row buffers, with other sizes (which would incur significant area and cost overheads) shown for reference.

## 5.1 Effects of Row Buffer Size on Energy

Reducing the size of the row buffer can reduce memory energy by requiring fewer components to buffer data. However, the reduction in row buffer locality may cause more accesses to the memory array, increasing memory energy. Figure 5a shows the main memory energy for a system with block interleaving and Figure 5d shows the main memory energy for a system with row interleaving.

In all cases, reducing the row buffer size can significantly reduce memory energy consumption, though there are diminishing marginal returns. The diminishing marginal returns are because, as the row buffer size decreases, memory energy becomes dominated by the energy required to transfer data between the row buffer and I/O pads during read and write operations.

At a row buffer size of 8B per chip, for example, data transfer comprises all of the memory energy. Thus, though row buffers can potentially prevent some device accesses, the energy required to keep the row buffers active may outweigh this benefit.

In addition, comparing Figures 5a and 5d, a row-interleaved address mapping consumes less energy than cache block-interleaved, but the difference between the two becomes smaller with smaller row buffer sizes. The smaller energy consumption can be attributed to the additional data accesses which are serviced in the row buffer at a lower latency and energy than accessing the memory array. The decreasing difference between the two memory mapping schemes is because as row buffer size becomes smaller, a row interleaved scheme has less row buffer space to exploit. At the smallest row buffer size of 8B per chip, block interleaving and row interleaving have the same memory energy consumption. Thus, though a block interleaving scheme consumes more energy than a row interleaving scheme on our system, smaller row buffers reduce the benefits of row interleaving, while greatly reducing energy.

We observe that, for an 8-core system, a modest row buffer size of 64B per chip leads to 47% and 67% less main memory dynamic energy consumption for PCM and STT-RAM, compared to an all-DRAM main memory with large rows (1KB per chip) for cache block interleaving and 28% and 52% less dynamic memory energy for row interleaving. Note that this reduction is achieved despite worse underlying technology metrics than DRAM (e.g., 2×/100× higher read/write energy per bit in PCM versus DRAM) because the energy saved by reducing the row buffer size more than makes up for the higher average memory access energy. Hence, an NVM main memory with smaller row buffers can significantly reduce memory energy consumption compared to a DRAM baseline with large row buffers.

## 5.2 Effects of Row Buffer Size on Performance

Requests can be serviced from the row buffer at a lower latency than accessing the memory device. However, a program's ability to access data from the row buffer depends on its memory access pattern, the row buffer size, system contention, and the memory mapping scheme. Based on the row buffer locality results from Figure 2 in Section 1, we hypothesize that the low row buffer locality present in the multi-core system we examine will not allow the available row buffer storage to be used effectively, regardless of the mapping scheme. We evaluate the performance of our system using the weighted speedup metric [23]: the sum of the speedups of the benchmarks when run together compared to when run alone on the same system with a DRAM main memory.

Figure 5b shows system performance under block interleaving and Figure 5e shows system performance under row interleaving. As the row buffer locality results from before suggested, for a given memory technology, reducing the row buffer size does not greatly affect system performance for either interleaving scheme due to the large amount of contention present on our multi-core system.

For the small performance degradation that we did observe, we find that decreasing the row buffer size reduces performance more for row interleaving than for cache block interleaving. This makes sense, because row interleaving works well with larger row buffer sizes: increasing row buffer size for access patterns which reference nearby addresses leads to more data accessed from the row buffer at a lower latency. Row buffer size has a very small effect on performance for cache block interleaving due to its low row buffer locality on our multi-core system (less than 10%, even at the largest row buffer size of 1KB).

Interestingly, with similar technology-dependent timing parameters as DRAM, an STT-RAM main memory can achieve slightly better performance. There are two reasons for this: (1) eliminating the precharge delay in NVMs leads to reduced latencies for row buffer misses, and (2) the relaxed $t_{RRD}$ and $t_{FAW}$ timing parameters due to reduced peak power and smaller row buffer sizes enable more banks to be accessed simultaneously in STT-RAM, enabling more memory system parallelism.

## 5.3 Effects of Row Buffer Size on Lifetime

NVM cells have a limited lifetime in terms of the number of times they may be written to before their ability to store data fails. We examine the effects of different row buffer sizes on device lifetime with and without a small e-DRAM cache before a PCM main memory, as an example of one technique to reduce writes to the main memory already employed in some processors [22]. The cache line size was the same as the other on-chip caches, 64B, and the cache was 8-way set associative, with a *Least Recently Used (LRU)* replacement policy. Figure 5c shows the normalized number of writes to the PCM main memory for systems with and without a 32MB e-DRAM cache for the row and block interleaving schemes.

We find that with or without a cache, decreasing the row buffer size has only a small effect on the number of NVM writes performed across both of the interleaving schemes. We find that writes, in fact, generally decrease slightly going to smaller row buffer sizes because the systems with smaller row buffers generally run slower (cf. PCM in Figures 5b and 5e), causing more requests to become delayed in the memory system for longer amounts of time, leading to more writes being coalesced than on the large row buffer systems which run faster.

We also find that the addition of a reasonably-sized e-DRAM cache (similar to that employed in existing systems [22]) has a large impact on the reduction of writes, decreasing the number of writes from between 39% to 47% across the various row buffer sizes.

Figure 5f shows the lifetime of the NVM main memory in years, were it to run the same workloads non-stop, for a PCM capacity of 2GB, a conservative cell endurance of $10^8$ writes [18, 11], and an ideal wear-leveling scheme which perfectly levels wear. The decrease in device lifetime with increasing row buffer sizes can be explained by the fact that systems with larger row buffers generally exhibit higher performance and thus issue more writes in the same amount of time as the smaller row buffer systems. We find that all configurations can support a lifetime of at least five years, with some exceeding ten years, making NVM memories of small row sizes feasible for use in many-core CMP systems.

# 6 Related Work

To our knowledge, we are the first to propose an NVM architecture and protocol to enable a single data path for reads and writes, and evaluate the system-level tradeoffs of row buffer sizes in NVM main memories for many-core CMP systems. Several related works have looked at the reorganization of the row buffer architecture in memory devices.

Udipi et al. proposed reorganizing the DRAM chip micro-architecture to address large buffer-induced energy waste [26]. The difficulty of performing this restructuring in DRAM lies in the fact that it either requires fine-grained array partitioning which incurs significant area overhead (their SBA technique) or requires increased design complexity with additional changes to the existing interface to the memory controller (their SSA technique). Our NVM architecture, on the other hand, does not require significant area penalty nor memory controller interface changes.

Lee et al. looked at employing multiple, more narrow rows in a PCM main memory for reducing the number of array reads and writes—and thus dynamic energy—without greatly reducing performance [11]. However, this past work employed the same architecture as existing DRAM designs and focused on an iso-area reorganization of row buffers and consequently would require more area overhead than our technique. In addition, it assumed a standard DRAM protocol for device access.

Multi-Core DIMM [3] and Mini-Rank [29] are techniques that partition a rank into smaller groups and allow the memory controller to issue commands to each group independently. As a result, multiple requests can be serviced in parallel, albeit at higher transfer latencies. The parallelism exposed by these techniques could be applied to our NVM architecture to potentially improve performance.

Hsu and Smith [10] employed multiple row buffers in the context of vector supercomputers and Hidaka et al. [9] described a DRAM chip with a small SRAM cache, however, their goal was not energy efficiency but performance, and their techniques incur significant area cost.

Micro-Pages [24] is a technique which can be used to improve row buffer locality and energy efficiency by aggregating frequently-accessed data in the same row. This technique could be combined with our NVM architecture to improve the row buffer locality of small row buffers.

# 7 Conclusions

To reduce the energy wasted by large row buffers in multi-core CMPs, we proposed a new organization and access protocol for NVMs that transfers and buffers only small amounts of data from the array to improve energy efficiency.

Using energy and timing models, we evaluated the system-level effects of our architecture in NVM main memories for many-core CMPs in terms of energy, performance, and durability and found that: (1) NVM main memories can achieve significant energy gains with reduced row buffer size (up to 67% less energy compared to DRAM). (2) These energy improvements can be achieved without significantly affecting performance. In fact, an STT-RAM-based main memory with small row buffers can achieve slightly better performance and much better energy-efficiency than DRAM due to the relaxed timing constraints of the NVM access protocol. (3) Even on a many-core CMP system with small row buffers, an NVM main memory is able to achieve a reasonable five year lifetime, with over ten years achievable with the addition of a small, 32MB e-DRAM cache.

# Acknowledgments

We thank the anonymous reviewers for their feedback. We gratefully acknowledge members of the SAFARI research group, CALCM, and for many insightful discussions on this work. This research was partially supported by an NSF

CAREER Award CCF-0953246, NSF Grant CCF-1147397, Gigascale Systems Research Center, Intel Corporation ARO Memory Hierarchy Program, and Carnegie Mellon CyLab. We also acknowledge equipment and gift support from Intel and Samsung.